\documentclass[showpacs,preprintnumbers,amsmath,amssymb,prb]{revtex4}

\usepackage[dvips]{graphicx}

\begin{document}
\title{Non-invertible Transformations and Spatiotemporal Randomness}

\author{J. A. Gonz\'alez}

\affiliation{Centro de F\'{\i}sica, Instituto Venezolano de Investigaciones
Cient\'{\i}ficas\\
Apartado 21827, Caracas 1020-A, Venezuela}

\author{A. J. Moreno and L. E. Guerrero}

\affiliation{Departamento de F\'{\i}sica, Universidad Sim\'on Bol\'{\i}var\\
Apartado Postal 89000, Caracas 1080-A, Venezuela}

\begin{abstract}
We generalize the exact solution to the Bernoulli shift map. Under
certain conditions, the generalized functions can produce unpredictable
dynamics. We use the properties of the generalized functions to show
that certain dynamical systems can generate random dynamics.
For instance, the chaotic Chua's circuit coupled to a circuit with a
non-invertible I-V characteristic can generate unpredictable dynamics.
In general, a nonperiodic time-series with truncated exponential 
behavior can be converted into unpredictable dynamics using 
non-invertible transformations.
Using a new theoretical framework for chaos and randomness, we 
investigate some classes of coupled map lattices. We show that,
in some cases, these systems can produce completely unpredictable
dynamics. In a similar fashion, we explain why some wellknown
spatiotemporal systems have been found to produce very complex
dynamics in numerical simulations. We discuss real physical
systems that can generate random dynamics.

\end{abstract}

\pacs{05.45.-a, 05.40.-a, 47.52.+j}
\date{\today}
\maketitle

\section{Introduction}

In the last decades, the strands of chaos theory have spread across all the
sciences like a fractal tree. Chaos theory and nonlinear dynamics have
provided new theoretical tools that allow us to understand the complex
behaviors of many physical systems
[Lorenz, 1993], [Schuster,1995], [Jackson, 1991], [Moon, 1991], [Strogatz, 1994], [Glass, 1998]. 

Deterministic chaotic behavior often looks erratic and random like the
behavior of a system perturbed by external noise. However, the known chaotic
systems are not random: precise knowledge of the initial conditions of the
system allows us to predict exactly the future behavior of that system, at
least in the short term.

In chaotic systems we can observe the divergence of nearby trajectories
[Lorenz, 1993], [Schuster,1995], [Jackson, 1991], [Moon, 1991], [Strogatz, 1994], [Glass \& Mackey, 1998].
This property represents a
difference between complex behavior due to deterministic chaos and that due to
true randomness [Lorenz, 1993].

This divergence of nearby trajectories leads to a kind of long-term
unpredictability. In the random systems we observe immediate unpredictability.
Already the next value is unpredictable.

There are processes, as the breaking sea waves on the shore, that are
deterministic although they seem random. The behavior of these processes is
determined by precise laws.

According to many definitions of randomness, in a random sequence of values,
the next value can be any of the previous values with equal probability
[Lorenz, 1993]. An example is the coin tossing experiments. Knowing the result
of the last coin tossing realization does not increase our chance to guess the
result of the next realization.

According to less strict definitions, in a random sequence, the next value can
be any of the possible values even if they possess different probabilities,
and even if their probability depends on the previous values [Lorenz, 1993]. In
other words, for the next outcome there is always more than one possible value.

On the other hand, in a non-random sequence, the next value is always
determined by the previous values
[Lorenz, 1993], [Schuster,1995], [Jackson, 1991], [Moon, 1991], [Strogatz, 1994], [Glass \& Mackey, 1998].

Can we explain all the randomness we observe in nature using the known
temporal chaotic systems?

Another very active area nowadays in nonlinear dynamics is spatiotemporal
chaos.

There are several paradigms and model equations for the studying of
spatiotemporal and extended systems
[Kaneko, 1985], [Kaneko, 1992], [Kaneko, 1989], [Chat\'{e} \& Manneville, 1988], [Crutchfield \& Kaneko, 1988], [Mayer-Kress \& Kaneko, 1989],
[Kuramoto, 1984], [Politi \& Torcini, 1992], [Kaneko, 1990], [Hansel \& Sompolinsky, 1993], [Bauer \textit{et al.}, 1993], [Bunimovich, 1995],
[Grassberger \& Scheiber, 1991], [Kaneko \& Konishi, 1989], [Kaneko, 1990a], [Chat\'{e}, 1995], [Gonz\'{a}lez \textit{et al.}, 1996],
[Gonz\'{a}lez \textit{et al.}, 1998], [Guerrero \textit{et al.}, 1999], [Pikovsky \& Kurths, 1994], [Bohr \textit{et al.}, 2001],
[Grigoriev, 1997], [Grigoriev \& Schuster, 1998], [Shibata \& Kaneko, 1998], [Wackerbauer \& Showalter, 2003], [Willeboordse, 2003], [Kaneko \& Tsuda, 2003].

Coupled map lattices are among the youngest models of extended dynamical
systems.

There is a vast literature dedicated to coupled iterated maps
[Kaneko, 1985], [Kaneko, 1992], [Kaneko, 1989], [Chat\'{e} \& Manneville, 1988], [Crutchfield \& Kaneko, 1988], [Mayer-Kress \& Kaneko, 1989],
[Grigoriev \& Schuster, 1998], [Shibata \& Kaneko, 1998], [Kaneko \& Tsuda, 2003].
Many important numerical results have been obtained in this area. However,
the behavior of such coupled systems is quite complex and by no means fully
explored.

Are there fundamental differences between the dynamics generated by temporal
and spatiotemporal systems?

Researchers have found [Chat\'{e}, 1995] that the usual temporal chaos methods of
time-series analysis are doomed when the dimension of the spatiotemporal
system becomes large (say larger than 10).

On the other hand, it is generally recognized
[Kaneko, 1985], [Kaneko, 1992], [Kaneko, 1989], [Chat\'{e} \& Manneville, 1988], [Crutchfield \& Kaneko, 1988], [Mayer-Kress \& Kaneko, 1989],
[Kuramoto, 1984], [Politi \& Torcini, 1992], [Kaneko, 1990], [Hansel \& Sompolinsky, 1993], [Bauer \textit{et al.}, 1993], [Bunimovich, 1995],
[Grassberger \& Scheiber, 1991], [Kaneko \& Konishi, 1989], [Kaneko, 1990a], [Chat\'{e}, 1995], [Gonz\'{a}lez \textit{et al.}, 1996],
[Gonz\'{a}lez \textit{et al.}, 1998], [Guerrero \textit{et al.}, 1999], [Pikovsky \& Kurths, 1994], [Bohr \textit{et al.}, 2001],
[Grigoriev, 1997], [Grigoriev \& Schuster, 1998], [Shibata \& Kaneko, 1998], [Wackerbauer \& Showalter, 2003], [Willeboordse, 2003], [Kaneko \& Tsuda, 2003]
that the dynamics of coupled maps is still far from being understood.

Cellular automata conform another class of dynamical systems that has been
studied intensively during the last years as simple models for spatially
extended systems. In this case, one replaces the continuous variables at each
space-time point by discrete ones
[von Neumann \& Burks, 1996], [Wolfram, 1983], [Wolfram, 1984], [Wolfram, 1984a], [Wolfram, 1986], 
[Hastings \textit{et al.}, 2003], [Israeli \& Goldenfeld, 2004].

In spite of their simplicity, automaton models are capable of describing many
features of physical processes
[Wolfram, 1983], [Wolfram, 1984], [Wolfram, 1984a], [Wolfram, 1986], 
[Hastings \textit{et al.}, 2003], [Israeli \& Goldenfeld, 2004].

Most results in the field of spatiotemporal systems have been obtained by
numerical simulations [Grigoriev \& Schuster, 1998].

In the present paper we will show that there exist dynamical systems that can
generate completely unpredictable dynamics in the sense that given any string
of generated values, for the next outcome, there is always more than one
possible value.

The mechanism responsible for the generation of randomness, in a very general
class of models and physical systems, is the presence of non-invertible
transformations of time-series that contain (truncated) exponential dynamics
or chaotic dynamics.

Using a new theoretical framework for randomness we will investigate some
classes of coupled map lattices. We will show, that in some cases, these
systems can produce completely unpredictable dynamics.

In a similar fashion, we will explain why some elementary cellular automata
with very simple rules have been found to produce very complex dynamics in
numerical simulations [Wolfram, 1986].

Some consequences of these results in the study of physical and economic
systems are discussed.

Some of the concepts discussed in the present paper about the differences
between common chaotic and random systems are inspired in
[Brown \& Chua, 1996] and [Gonz\'{a}lez \textit{et al.}, 2000].

\section{Unpredictable Dynamics}

We will call a time-series $\left\{  X_{n}\right\}  $ unpredictable if for any
string of $m+1$ numbers $X_{0},X_{1},X_{2},...,X_{m}$ ($m$ can be as large as
we wish), then the next number $X_{m+1}$ can take more than one value.

Let us define the general function
\begin{equation}
X_{n}=P\left(  \theta Tz^{n}\right)  , \label{1}%
\end{equation}
where $P\left(  t\right)  $ is a periodic function, $T$ is the period of
function $P\left(  t\right)  $, $\theta$ and $z$ are real numbers.
An
important example of function $P\left(  t\right)  $ is function $P\left(
t\right)  =t\pmod 1 $. Note that this is a periodic function with period $T=1$:
$P\left(  t+1\right)  =\left(  t+1\right)  \pmod 1 =P\left(  t\right)  $.

We will show that the dynamics contained in function (\ref{1}) is unpredictable.

Let us define the family of sequences
\begin{equation}
X_{n}^{(k,m)}=P\left[  T\left(  \theta_{0}+q^{m}k\right)  z^{n}\right]  ,
\label{2}%
\end{equation}
where $z=p/q$ is a rational number such that $p$ and $q$ are relative primes
($p>q$), $k$ and $m$ are integers. The parameter $k$ distinguishes the
different sequences. For all sequences parametrized by $k$, the first $m+1$
values are the same. This is true because
\begin{equation}
X_{n}^{(k,m)}=P\left[  T\theta_{0}\left(  p/q\right)  ^{n}+Tkp^{n}%
q^{(m-n)}\right]  =P\left[  T\theta_{0}\left(  p/q\right)  ^{n}\right]  ,
\label{3}%
\end{equation}
for all $n\leq m$. Note that the number $kp^{n}q^{(m-n)}$ is an integer for
$n\leq m$.

The interesting conclusion is that the next value
\begin{equation}
X_{m+1}^{(k,m)}=P\left[  T\theta_{0}\left(  p/q\right)  ^{m+1}+Tkp^{(m+1)}%
/q\right]  \label{4}%
\end{equation}
is unpredictable. $X_{m+1}^{(k,m)}$ can take $q$ different values. For a
generic real $z$, $X_{m+1}$ can take an infinite number of values.

Let us discuss some properties of the following particular case of function
(\ref{1}):
\begin{equation}
X_{n}=\theta z^{n}\pmod 1  . \label{5}%
\end{equation}

For $z=2$, function (\ref{5}) is the exact solution to the Bernoulli shift map.

\begin{figure}[ptb]
\centerline{\includegraphics[width=5in]{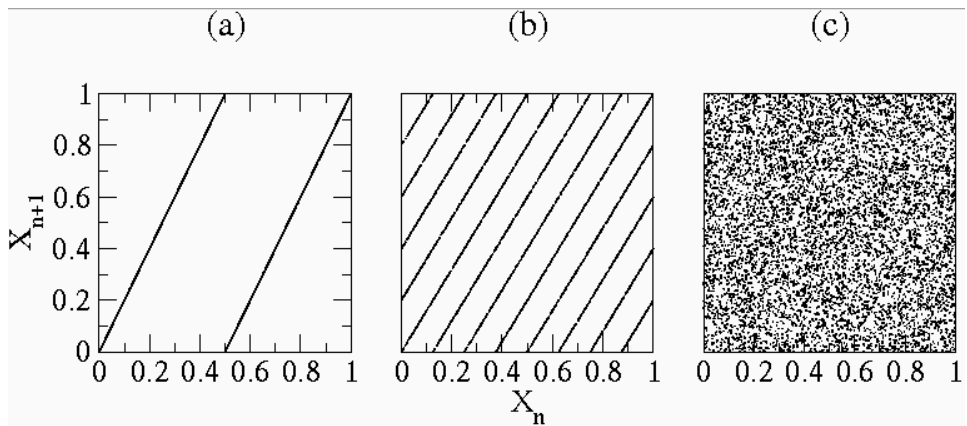}}
\caption{First-return maps constructed using the dynamics produced
by function (\ref{5}) for
$\theta=\pi$. (a) $z=2$. (b) $z=8/5$. (c) $z=\pi$.}
\label{fig1}
\end{figure}

Figures \ref{fig1}(a)-(c) show different examples of first-return maps
produced by the dynamics represented by Eq. (\ref{5}). The values generated by
function (\ref{5}) are uniformly distributed in the interval $0<X_{n}<1$.

We have generalized these results to functions of type
\begin{equation}
X_{n}=h\left[  f\left(  n\right)  \right]  . \label{6}%
\end{equation}

To produce complex dynamics, the function $f(n)$ does not have to be
exponential all the time, and function $h(y)$ does not have to be 
periodic [Gonz\'{a}lez \textit{et al.}, 2002]. In
fact, it is sufficient for function $f(n)$ to be a finite nonperiodic
oscillating function which possesses repeating intervals of truncated
exponential behavior. For instance, this can be a common chaotic sequence.

On the other hand, function $h(y)$ should be non-invertible. In other words,
it should have different maxima and minima in such a way that equation
$h(y)=\alpha$ (for some specific interval of $\alpha$, $\alpha_{1}%
<\alpha<\alpha_{2}$) possesses several solutions for $y$. Of course, the image
of function $f(n)$ should be in the interval where function $h(y)$ is noninverible.

\begin{figure}[ptb]
\centerline{\includegraphics[width=5in]{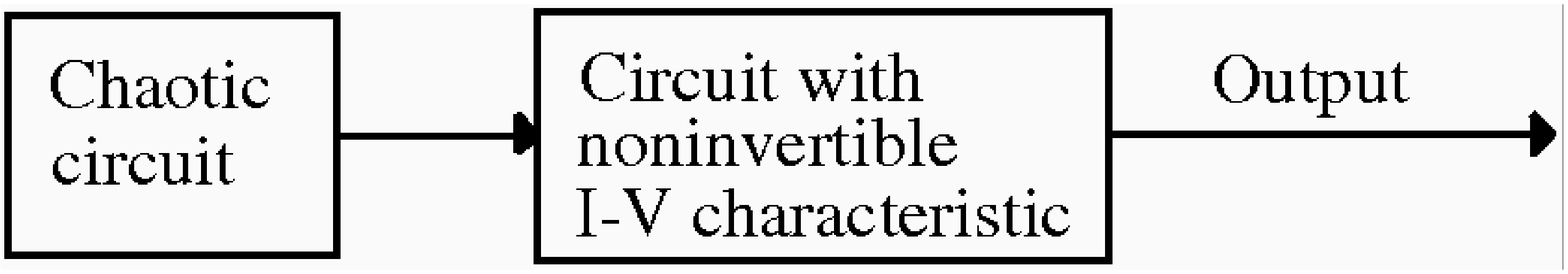}}
\caption{Scheme of a system that can produce unpredictable dynamics.}
\label{fig2}
\end{figure}

Gonz\'{a}lez \textit{et al.} [2002] have shown that a chaotic Chua's circuit
[Matsumoto \textit{et al.}, 1985], [Matsumoto \textit{et al.},1987]
coupled to a Josephson junction can generate
unpredictable dynamics. In fact, in order to produce unpredictable dynamics we
can use a system with the features shown in Fig. \ref{fig2}.

A method for the construction of circuits with non-invertible $I-V$
characteristics can be found in
[Chua \textit{et al.}, 1987] and [Comte \& Marqui\'{e}, 2002].

\section{Finite Systems of Coupled Maps}

Let us consider the following dynamical system
\begin{equation}
X_{n+1}=\begin{cases}aX_n, & \text{if $X_n<Q$,}\\
bY_n, & \text{if $X_n>Q$,}  
\end{cases}
\label{7}%
\end{equation}%
\begin{equation}
Y_{n+1}=cZ_{n}, \label{8}%
\end{equation}%
\begin{equation}
Z_{n+1}=X_{n}\pmod 1  . \label{9}%
\end{equation}

Here $a$ can be an irrational number, $a>1$, $b>1$, $c>1$. We can note that
for $0<X_{n}<Q$, the behavior of function $Z_{n}$ is exactly like that of
function (\ref{5}). For $X_{n}>Q$ the dynamics is re-injected to the region
$0<X_{n}<Q$\ with a new initial condition. While $X_{n}$ is in the interval
$0<X_{n}<Q$, the dynamics of $Z_{n}$ is unpredictable as it is function
(\ref{5}). Thus, the proccess of producing a new initial condition through Eq.
(\ref{8}) is random.

If the only observable is $Z_{n}$, then it is impossible to predict the next
values of this sequence using only the knowledge of the past values of
$\left\{  Z_{n}\right\}  $.

\begin{figure}[ptb]
\centerline{\includegraphics[width=5in]{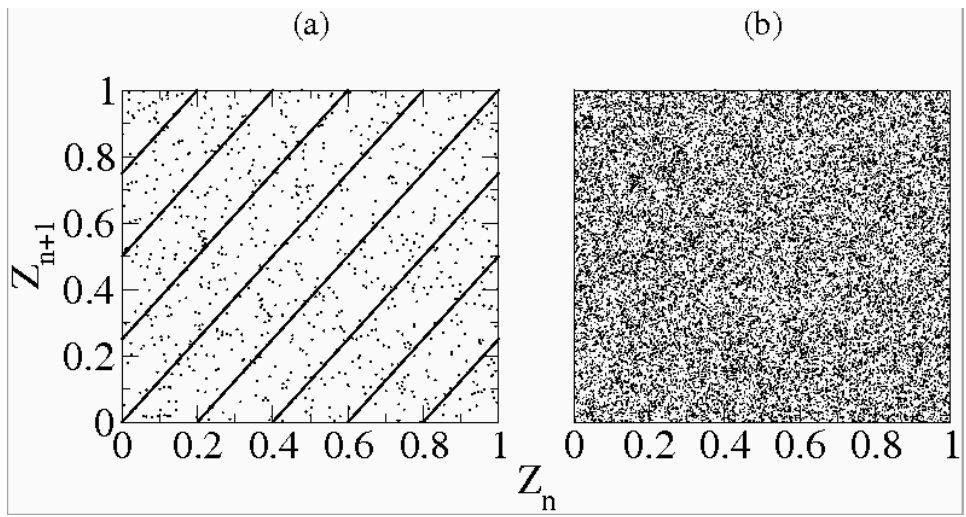}}
\caption{First-return maps produced by dynamical system (\ref{7})-(\ref{9}) for
$X_{0}=Y_{0}=Z_{0}=0.1$, $Q=200$, $b=c=2$. (a) $a=5/4$. (b) $a=\pi$.}
\label{fig3}
\end{figure}

An example of the dynamics produced by the dynamical system (\ref{7}%
)-(\ref{9}) is shown in Fig. \ref{fig3}.

In the dynamical system (\ref{7})-(\ref{9}) the variable $Z_{n}$ is quasi-random,
but the variable $X_{n}$ is predictable because in the interval $0<X_{n}<Q$
the rule to determine the next number is a one-valued function.

In principle, we can construct dynamical systems where all the variables
(taken separately) are random.

Consider the following system:
\begin{equation}
X_{n+1}=\begin{cases}\left( a+bZ_{n}\right) X_{n}+cY_{n},%
& \text{if $X_n<Q$,}\\
bY_{n}, & \text{if $X_n>Q$,}  
\end{cases}
\label{10}%
\end{equation}%
\begin{equation}
Y_{n+1}=cZ_{n}, \label{11}%
\end{equation}%
\begin{equation}
Z_{n+1}=X_{n}\pmod 1  . \label{12}%
\end{equation}

Note that $X_{n}$, in Eq. (\ref{10}), still possesses a finite exponential
behavior for $0<X_{n}<Q$, because $\left(  a+bZ_{n}\right)  $ is always a
positive number. However, in this case the dynamics of $X_{n}$ is influenced
all the time by the random dynamics of $Z_{n}$.

If we are interested in dynamical systems where all the variables are random
and uniformly distributed in the interval $\left[  0,1\right]  $, then we can
use the following one:
\begin{equation}
X_{n+1}=\left[  \left(  a+bZ_{n}\right)  X_{n}+cY_{n}+0.1\right]  \pmod 1  ,
\label{13}%
\end{equation}%
\begin{equation}
Y_{n+1}=\left[  dZ_{n}+fX_{n}+0.1\right]  \pmod 1  , \label{14}%
\end{equation}%
\begin{equation}
Z_{n+1}=\left[  gX_{n}+0.1\right]  \pmod 1  . \label{15}%
\end{equation}

Here $X_{n}$\ shares many of the properties that are present in the system
(\ref{10})-(\ref{12}).

\begin{figure}[ptb]
\centerline{\includegraphics[width=5in]{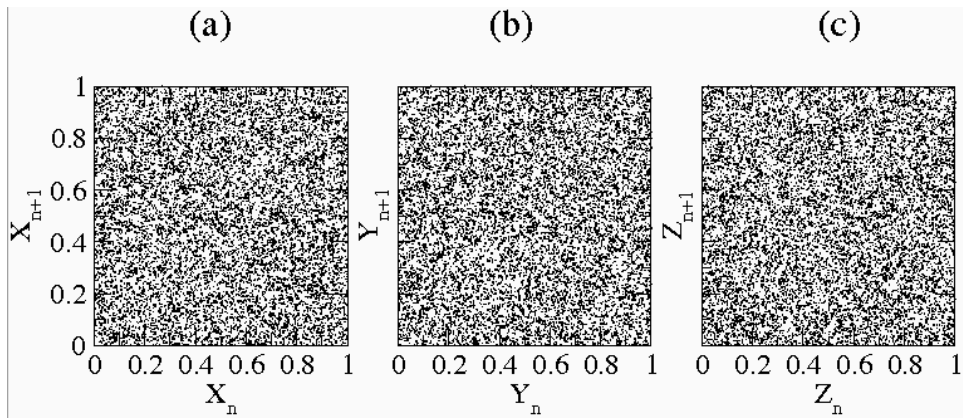}}
\caption{Typical dynamics generated by dynamical system (\ref{13})-(\ref{15}).
All the variables are unpredictable. Parameter values: $a=4/3$, $b=2.1$, 
$c=7.3$, $d=3.1$, $f=7.7$, $g=113$.
Initial conditions: $X_{0}=Y_{0}=Z_{0}=0.1$. (a) First-return map of variable
$X_{n}$. (b) The same for variable $Y_{n}$. (b) The same for variable $Z_{n}$.} 
\label{fig4}
\end{figure}

First-return maps of the time-series produced by dynamical system
(\ref{13})-(\ref{15}) can be observed in Fig. \ref{fig4}.

\section{Coupled Map Lattices}

Suppose now that we are interested in symmetric equations in the sense that
all equations for $X_{n}$, $Y_{n}$ and $Z_{n}$ are equivalent.

Note that in the dynamical systems (\ref{7})-(\ref{9}), (\ref{10})-(\ref{12})
and (\ref{13})-(\ref{15}), the equation for $Z_{n+1}$ is constructed in such a
way that a nonperiodic dynamics with truncated exponential behavior is the
argument of a non-invertible function (say $y=x\pmod 1   $). What function is
in the argument is not so important. So we can use a function that depends on
$X_{n}$, but also on $Y_{n}$ and $Z_{n}$ as well. On the other hand, the most
important feature of the equation for $Y_{n+1}$\ is that it depends on the
random variable $Z_{n}$. So it can also depend on $X_{n}$ and $Y_{n}$. Thus,
let us transform dynamical systems (\ref{13})-(\ref{15}) into a symmetric
system:
\begin{equation}
X_{n+1}=\left[  \left(  a_{1}+b_{1}Y_{n}+c_{1}Z_{n}\right)  X_{n}+d_{1}%
Y_{n}+e_{1}Z_{n}\right]  \pmod 1  , \label{16}%
\end{equation}%
\begin{equation}
Y_{n+1}=\left[  \left(  a_{2}+b_{2}Z_{n}+c_{2}X_{n}\right)  Y_{n}+d_{2}%
Z_{n}+e_{2}X_{n}\right]  \pmod 1  , \label{17}%
\end{equation}%
\begin{equation}
Z_{n+1}=\left[  \left(  a_{3}+b_{3}X_{n}+c_{3}Y_{n}\right)  Z_{n}+d_{3}%
X_{n}+e_{3}Y_{n}\right]  \pmod 1  . \label{18}%
\end{equation}

\begin{figure}[ptb]
\centerline{\includegraphics[width=5in]{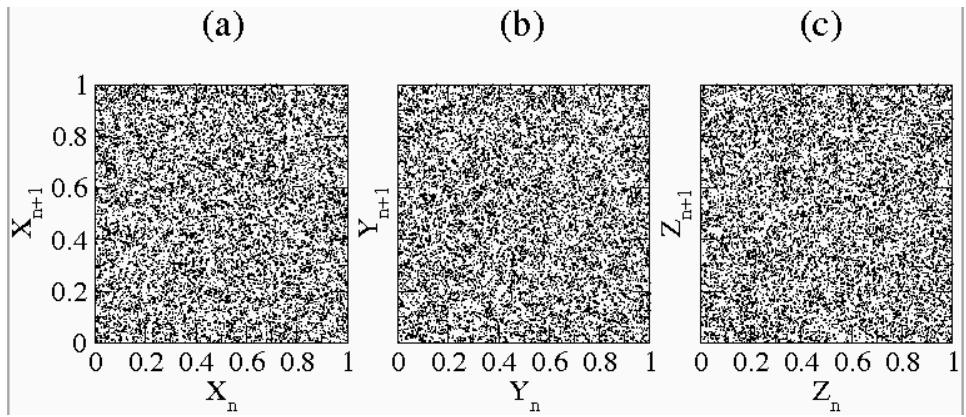}}
\caption{Dynamics produced by the set of Eqs. (\ref{16})-(\ref{18}).
Parameter values: $a_{1}=1.3$, $b_{1}=\pi$, $c_{1}=2.6$, $d_{1}=1.5$,
$e_{1}=1.1$, $a_{2}=4.6$, $b_{2}=2.1$, $c_{2}=e$, $d_{2}=3.2$,
$e_{2}=7.1$, $a_{3}=2.9$, $b_{3}=5.4$, $c_{3}=8.7$, $d_{3}=4.5$, $e_{3}=1.9$.
Initial conditions: $X_{0}=Y_{0}=Z_{0}=0.1$. (a) First-return map of variable
$X_{n}$. (b) The same for variable $Y_{n}$. (b) The same for variable $Z_{n}$.}
\label{fig5}
\end{figure}

Like the systems discussed before, the set of Eqs. (\ref{16})-(\ref{18})
will produce unpredictable dynamics for all the variables $X_{n}$, $Y_{n}$ and
$Z_{n}$ taken separately. This can be seen in Figs. \ref{fig5}(a)-(c).

Can we construct a coupled map lattice with these characteristics?

Now we will have a dynamical variable that depends on the time $n$ and the
space coordinate $i$. Instead of three equations with three variables as in
system (\ref{16})-(\ref{18}), we will have an infinite number of equations.
Our variable will be $X_{n}(i)$.

An example of a coupled map lattice with all the properties discussed above is
the following:
\begin{equation}
X_{n+1}(i)=\left[  \left(  a+bX_{n}(i-1)+cX_{n}(i+1)\right)  X_{n}%
(i)+dX_{n}(i-1)+fX_{n}(i+1)+0.1\right]  \pmod 1 .\label{19}%
\end{equation}

Note that for each space site $i$, we have a nonperiodic dynamics with
truncated exponential behavior that depends on the behavior of the space sites
$\left(  i-1\right)  $ and $\left(  i+1\right)  $. This dynamics is always the
argument of a noninvertible function (in this case $y=x\pmod 1 $). We are sure
that the dynamics is nonperiodic because even something as simple as
$X_{n+1}(i)=aX_{n}(i)\pmod 1 $ would produce chaotic behavior for $a>1$.

Another interesting example of coupled map lattices with random behavior can be
found in the system
\begin{equation}
X_{n+1}(i)=\left[  \left(  a+bX_{n}(i-1)+cX_{n}(i)+dX_{n}(i-1)\right)
X_{n}(i)+fX_{n}(i-1)+gX_{n}(i+1)+0.1\right]  \pmod 1 .\label{20}%
\end{equation}

Here the coefficient of variable $X_{n}(i)$ in the argument of the modulo
function depends on $X_{n}(i-1)$, $X_{n}(i+1)$ and the same $X_{n}(i)$.

\begin{figure}[ptb]
\centerline{\includegraphics[width=5in]{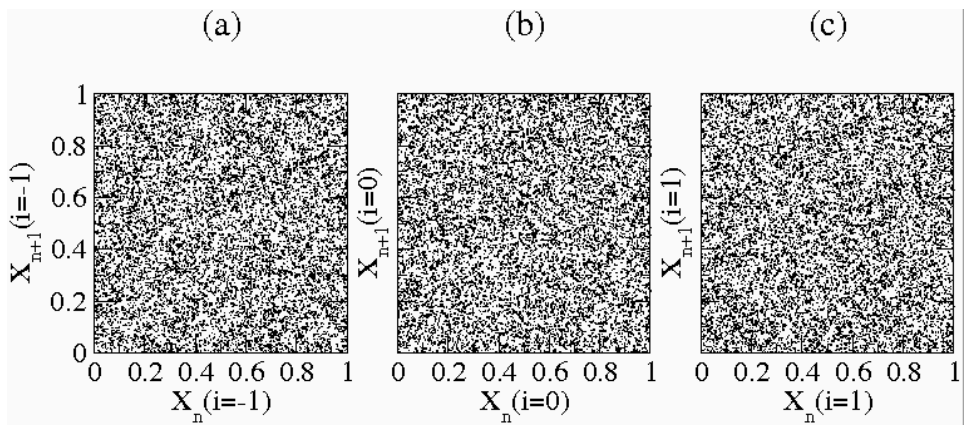}} \caption{Typical
dynamics generated by the coupled map lattice defined by Eq. (\ref{19}).
Parameter values: $a=2$, $b=c=d=f=1$. Initial condition: $X_{0}=0.1$.
(a) $i=-1$. (b) $i=0$. (c) $i=1$.}%
\label{fig6}%
\end{figure}

Figures \ref{fig6}(a)-(c) show the dynamics generated by different sites in
the introduced coupled map lattices.

\section{Cellular Automata}

The values of the sites in a one-dimensional cellular automaton are updated in
parallel in discrete time steps according to a rule of the form
\begin{equation}
Y_{n+1}(i)=F\left[  Y_{n}(i-r),Y_{n}(i-r+1),...,Y_{n}(i+r)\right]  .
\label{21}%
\end{equation}

The site values are usually taken as integers between zero and $\left(
k-1\right)  $ [Wolfram, 1983], [Wolfram, 1984], [Wolfram, 1984a], [Wolfram, 1986].

Cellular automata can be considered as discrete approximations to partial
differential equations, and used as direct models for a wide class of natural
systems [Wolfram, 1983], [Wolfram, 1984], [Wolfram, 1984a], [Wolfram, 1986],
[Hastings \textit{et al.}, 2003], [Israeli \& Goldenfeld, 2004].

A classification and several studies of the cellular automata with $k=2$ and
$r=1$ can be found in [Wolfram, 1983], [Wolfram, 1984], [Wolfram, 1984a], [Wolfram, 1986].

Representations of the so-called Rules 30, 110 and 124 are shown in
tables \ref{table1}-\ref{table3}.

\begin{table}[tbp]
\begin{center}
\begin{tabular}{|c|c|c|c|c|c|c|c|} \hline
111 & 110 & 101 & 100 & 011 & 010 & 001 & 000 \\ \hline
0 & 0 & 0 & 1 & 1 & 1 & 1 & 0 \\ \hline
\end{tabular}
\caption{Representation of Rule 30 cellular automaton.}
\label{table1}
\end{center}
\end{table}

\begin{table}[tbp]
\begin{center}
\begin{tabular}{|c|c|c|c|c|c|c|c|}
\hline
111 & 110 & 101 & 100 & 011 & 010 & 001 & 000 \\ \hline
0 & 1 & 1 & 0 & 1 & 1 & 1 & 0 \\ \hline
\end{tabular}
\caption{Representation of Rule 110 cellular automaton.}
\label{table2}
\end{center}
\end{table}

\begin{table}[tbp]
\begin{center}
\begin{tabular}{|c|c|c|c|c|c|c|c|}
\hline
111 & 110 & 101 & 100 & 011 & 010 & 001 & 000 \\ \hline
0 & 1 & 1 & 1 & 1 & 1 & 0 & 0 \\ \hline
\end{tabular}
\caption{Representation of Rule 124 cellular automaton.}
\label{table3}
\end{center}
\end{table}

The top row in each set of three elements gives one of the possible
combinations of values for a cell and its immediate neighbors. The bottom row
then specifies what value the center cell should have on the next step in each
of these cases.

Rules 110 and 124 are equivalent under reflection transformations [Wolfram, 1983].

Rules 110 and 124 are relevant because they have been proved to be equivalent to
Turing machines. So they are capable of universal computations [Israeli \& Goldenfeld, 2004].

On the other hand, Rule 30 has been considered as a model of randomness in
nature and has been used as a practical pseudorandom number generator
[Wolfram, 1986].

Rule 30 can be written as a coupled map lattice:
\begin{equation}
Y_{n+1}(i)=\left[  Y_{n}(i-1)+Y_{n}(i)+Y_{n}(i+1)+Y_{n}(i)Y_{n}(i+1)\right]
\pmod 2 .\label{22}%
\end{equation}

The sequences generated by Rule 30 have been analyzed by a variety of
empirical and statistical techniques [Wolfram, 1986] and the researchers have
concluded that they seem completely random.

Random sequences are obtained from Rule 30 by sampling the values that a
particular site attains as a function of time.

\begin{figure}[ptb]
\centerline{\includegraphics[width=5in]{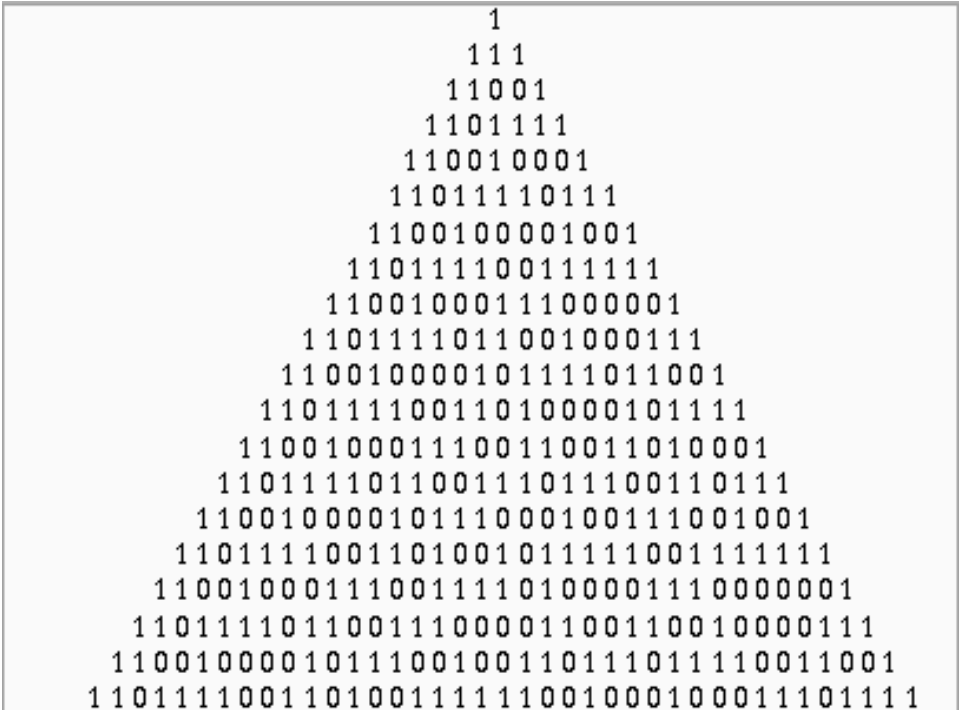}}
\caption{Evolution of Rule 30 cellular automaton.}
\label{fig7}
\end{figure}

\begin{figure}[ptb]
\centerline{\includegraphics[width=5in]{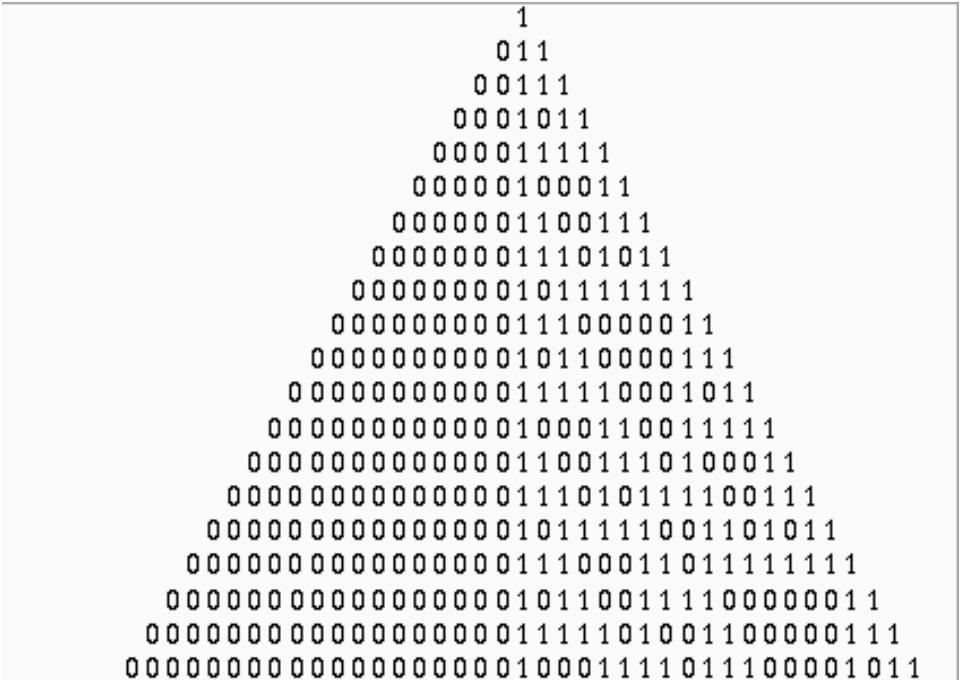}}
\caption{Evolution of Rule 124 cellular automaton.}
\label{fig8}
\end{figure}

An example very frequently used is the apparent ``randomness'' of the center
vertical column in the patterns shown in Fig. \ref{fig7}. The evolution
of Rule 124 cellular automaton is shown in Fig. \ref{fig8}.

In all these works the authors recognize that little has been proved
theoretically about Rule 30. However, the center vertical sequence has passed
all the statistical tests of randomness applied to it.

\begin{figure}[ptb]
\centerline{\includegraphics[width=5in]{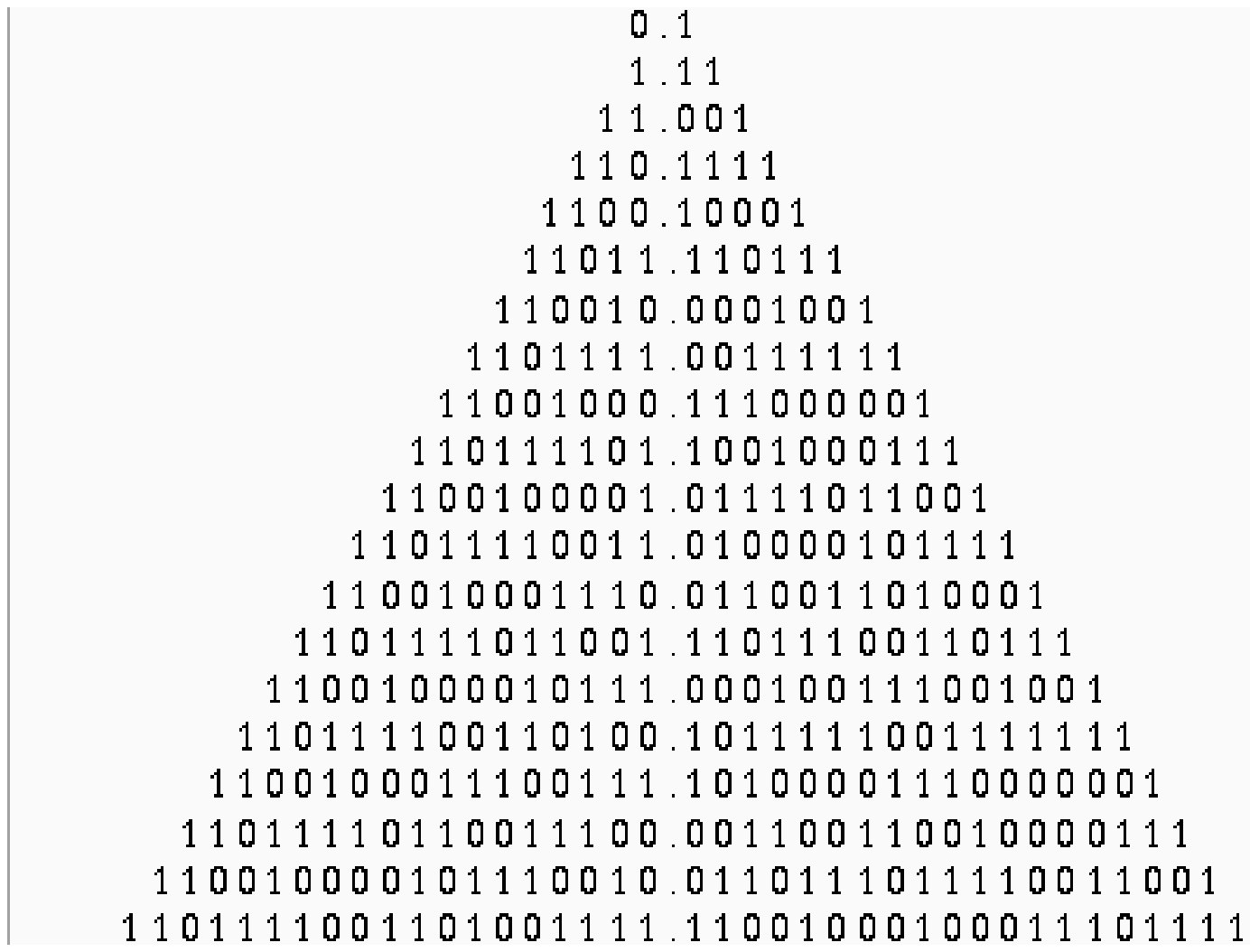}}
\caption{The outcomes of Rule 30 can be seen as a numerical time series of 
real numbers written in binary representation.}
\label{fig9}
\end{figure}

\begin{figure}[ptb]
\centerline{\includegraphics[width=5in]{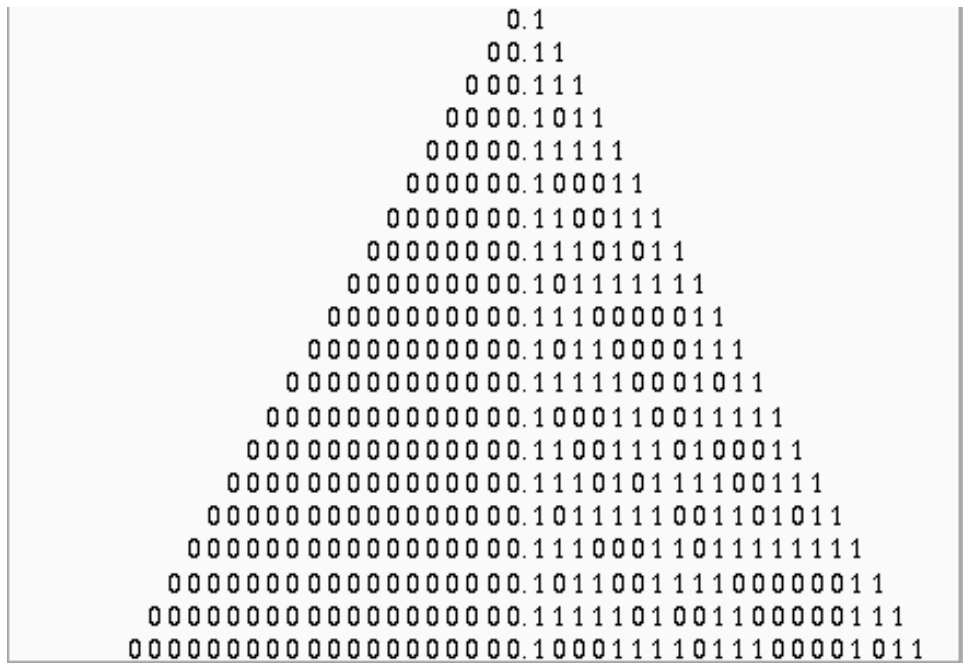}}
\caption{Binary representation of the time series produced by Rule 124.}
\label{fig10}
\end{figure}

If a point that separates the integer part from the fractionary part is placed
near the central column as is shown in Figures \ref{fig9}-\ref{fig10}, then the outcomes of the
cellular automaton evolution can be transformed into a numerical time series
$\left\{  Y_{n}\right\}  $, where the $Y_{n}$\ are real numbers written in
binary system.

In the case of Rule 124 this is always a bounded time series where $0\leq
Y_{n}\leq1$.

\begin{figure}[ptb]
\centerline{\includegraphics[width=5in]{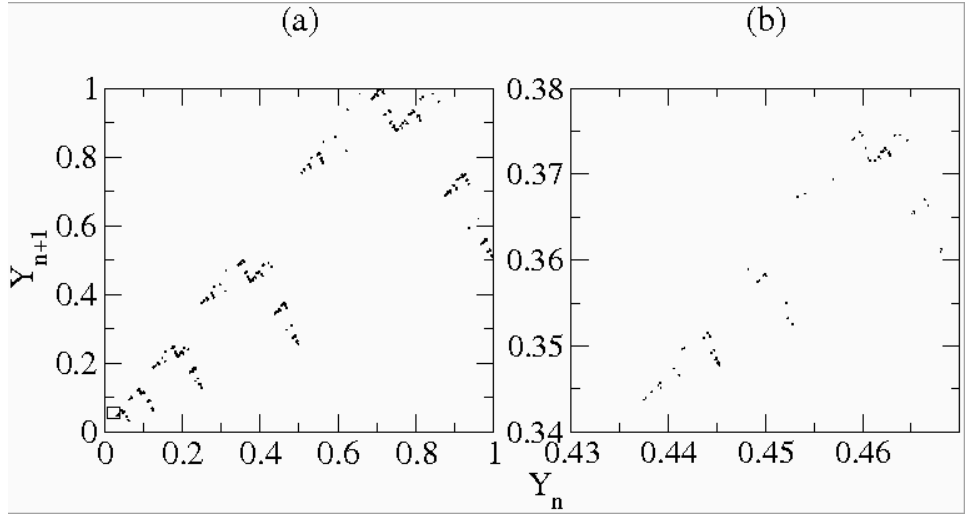}}
\caption{First-return map constructed using the sequence 
$\left\{  Y_{n}\right\}$
produced
with the dynamics of Rule 124 cellular automaton as described in the main
text. (a) Full first-return map. (b) Zoom of a detail of the first-return map.}
\label{fig11}
\end{figure}

The first-return map of a typical Rule 124 time series is shown in Fig.
\ref{fig11}. Note that the function $Y_{n+1}=f\left(  Y_{n}\right)  $ is a
fractal. Nevertheless, it is important to notice that despite its fractal
structure this is a one-valued first-return map. In fact, given any previous
value $Y_{n}$, the next value is always defined by this previous value. 
We
have calculated numerically the Lyapunov exponent of this map using different
methods of time-series analysis, see e.g. [Wolf  \textit{et al.}, 1985] and
[Kantz \& Schreiber, 1997]. In these calculations, we generate the sequence
$\left\{  Y_{n}\right\}  $ using the cellular automaton rule. Then we treat
the produced sequence as an experimental time-series, and we compute the
largest Lyapunov exponent using the mentioned standard methods. In our
calculations, the largest Lyapunov exponent is approximately
$\lambda=0.4$.
All this leads to the speculation
that a dynamical system, that can be mapped to a fractal chaotic map of
type\ $Y_{n+1}=f\left(  Y_{n}\right)  $, is capable of universal computation.

The geometrical structure shown in Fig. \ref{fig11} is an invariant and can be
used to have a general representation of the dynamics for any initial
condition. It is independent of time.

We believe that this kind of representation is more general and useful that
the Wolfram's ``space-time'' calculations of several hundreds or thousands
steps, because they are by definition limited and misleading. The dynamics
that can be observed in an interval of time of 1,000 steps can be very
different in an interval of time taken 1,000,000 steps away.

\begin{figure}[ptb]
\centerline{\includegraphics[width=5in]{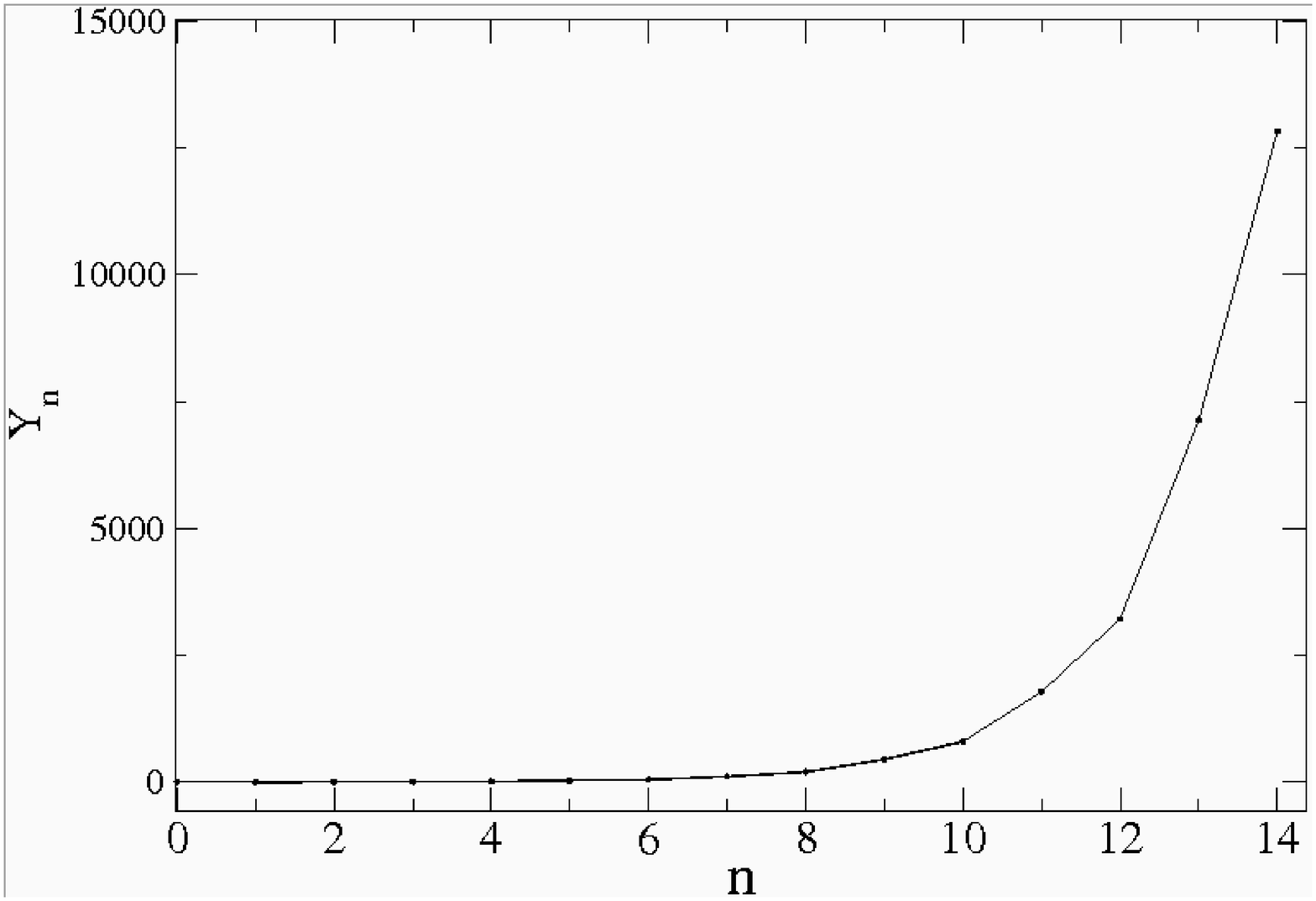}}
\caption{Approximate exponential behavior of $\left\{  Y_{n}\right\}$
for Rule 30.}
\label{fig12}
\end{figure}

On the other hand, for Rule 30, the time series $\left\{  Y_{n}\right\}  $ is
an unbounded exponentially increasing function (see Fig. \ref{fig12}). In fact,
$\left\{  Y_{n}\right\}  $ can be expressed as a map of type
\begin{equation}
Y_{n+1}=a_{n}Y_{n}, \label{23}%
\end{equation}
where $a_{n}$ always takes non-integer values such that $a_{n}>1$.

From the representation of Rule 30 in Table \ref{table1} it is evident that from a
number
\begin{equation}
Y_{n}=...b_{-3}b_{-2}b_{-1}.b_{1}b_{2}b_{3}b_{4}... \label{24}%
\end{equation}
where $b_{k}$, $b_{-k}$ are zeroes or ones; the number
\begin{equation}
Y_{n+1}=...b_{-3}^{\prime}b_{-2}^{\prime}b_{-1}^{\prime}.b_{1}^{\prime}%
b_{2}^{\prime}b_{3}^{\prime}b_{4}^{\prime}... \label{25}%
\end{equation}
can be obtained only using a non-integer $a_{n}$ which should be close to $2$
(see the actual evolution in Fig. \ref{9}). The behavior of $a_{n}$ can be seen
in Fig. \ref{13}.

\begin{figure}[ptb]
\centerline{\includegraphics[width=5in]{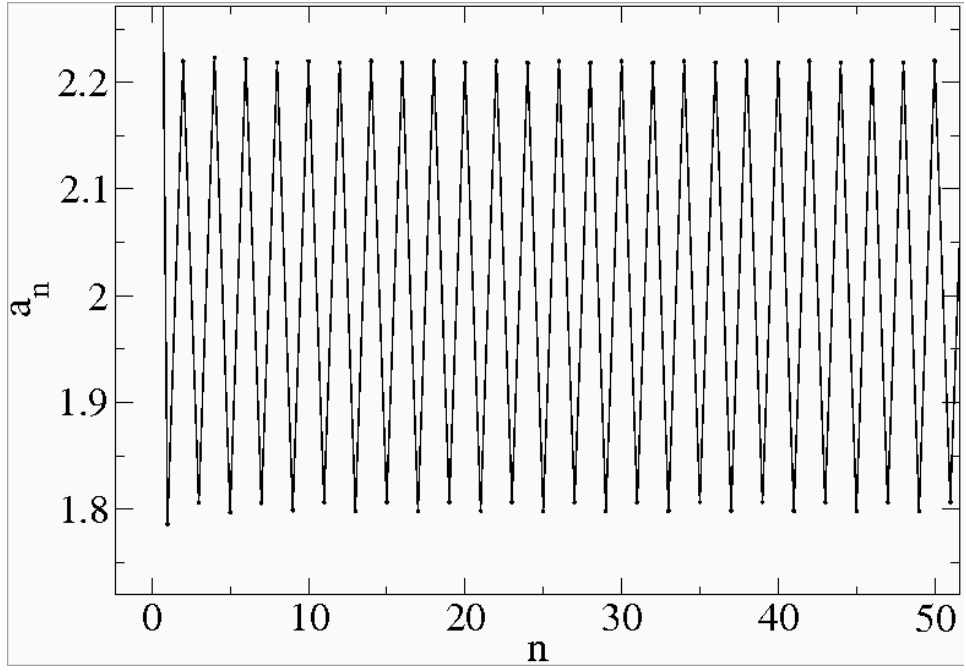}}
\caption{Behavior of $a_{n}$ as defined in Eq. (\ref{23}).}
\label{fig13}
\end{figure}

In fact, a numerical calculation shows that the dynamics of $\left\{
a_{n}\right\}  $\ possesses a quasiperiodic attractor (see Fig. \ref{fig13}) where
all the values of $a_{n}$ are close to two possible values: $1.8$ and $2.2$.

\begin{figure}[ptb]
\centerline{\includegraphics[width=5in]{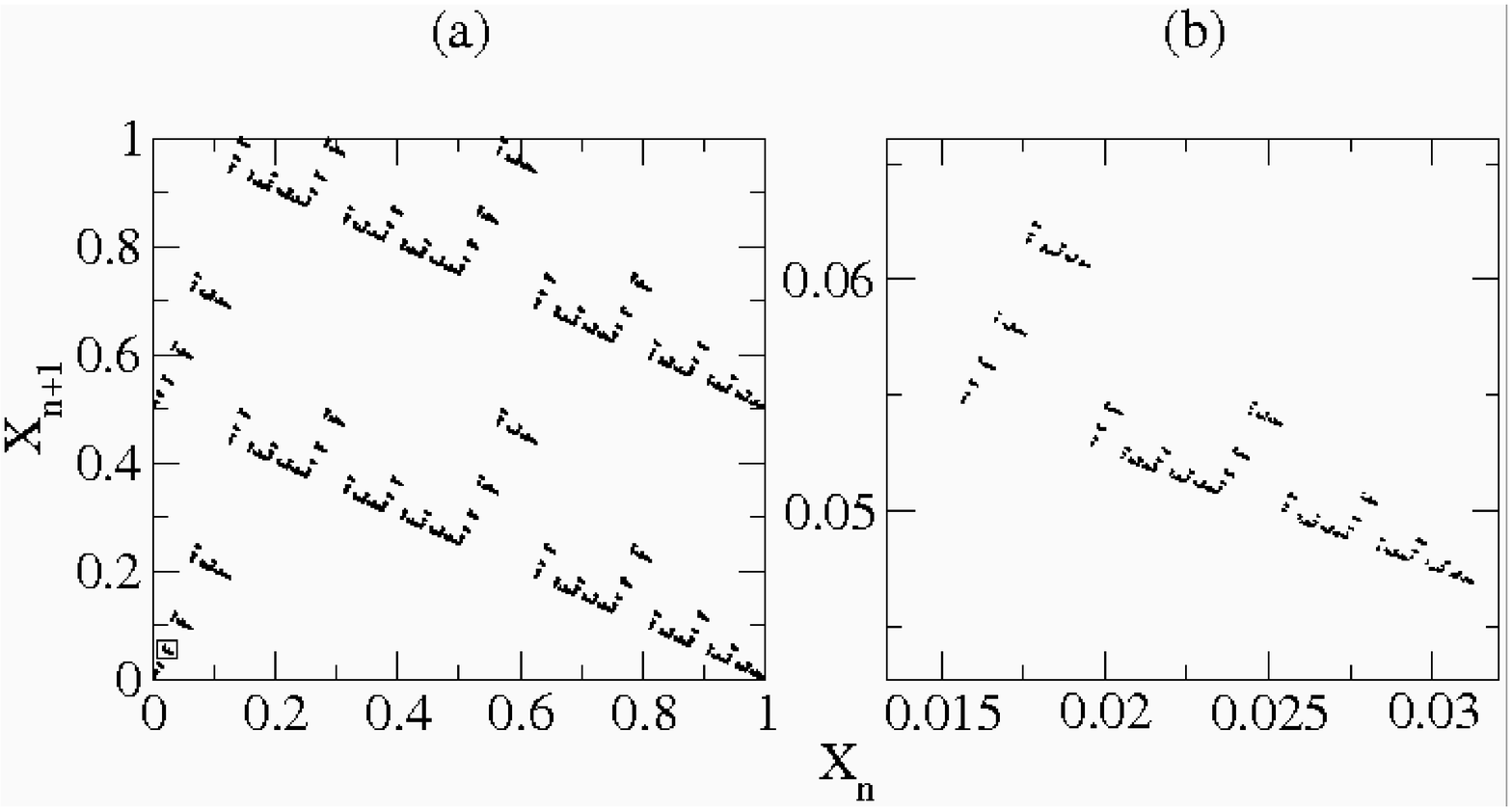}}
\caption{First-return map produced by Eq. (\ref{26}), where $Y_{n}$ is the
time-series generated by Rule 30 cellular automaton. (a) Full map. (b) Detail.}
\label{fig14}
\end{figure}

Thus, $Y_{n}$ is approximately an exponentially increasing function. All we need to produce
unpredictable dynamics is the application of a non-invertible transformation
on $Y_{n}$. For instance, the function
\begin{equation}
X_{n}=Y_{n}\pmod 1  \label{26}%
\end{equation}
is much more harder to predict than Rule 124. The first-return map of this
dynamics can be observed in Fig. \ref{fig14}. Note that the first-return map is
two-valued. Given a $X_{n}$, we always have two possible future values
$X_{n+1}$.

Notice that if $a_{n}=2$, the time-series is predictable as in the Bernoulli shift.

So it is important that, due to the structure of Rule 30, the resulting
$a_{n}$ are always non-integer. This leads to unpredictability.

Other non-invertible transformations can be used to produce unpredictability
from Rule 30.

The operation of sampling the value that a particular site attains as a
function of time is a non-invertible transformation. This can be observed in
table \ref{table1}, where Rule 30 is represented. So, by sampling the value of the
central column in the Rule 30 evolution (see Fig. \ref{fig9}), which is
equivalent to a sequence with exponential behavior, we are generating \ a much
\ more unpredictable dynamics than that produced by Rule 124.

All the studies about this cellular automaton conclude that its dynamics is
computationally as sophisticated as any physically realizable system
can be [Wolfram, 1983], [Wolfram, 1984], [Wolfram, 1984a], [Wolfram, 1986].

The authors of these works say that it is computationally irreducible and its
outcome can effectively be found only by direct simulation or observation. So
there should be no general computational shortcuts or finite mathematical
procedure to investigate its behavior. As a consequence, all the questions
concerning infinite time or infinite size limits cannot be answered by bounded computations.

In fact, they cannot be sure if after a very large number of time steps, the
dynamics generated by Rule 30 can become periodic.

However, following our results, we can predict that the dynamics produced by
sampling the values that a particular site attains as a function of time is nonperiodic.

From this we arrive at a very important conclusion concerning forecasting
methods in distributed systems. Although the general spatiotemporal dynamics
can be deterministic, the local dynamics can seem completely random.

In other words, chaotic spatiotemporal systems can produce completely random
dynamics locally. However, a knowledge of system spatiotemporal dynamics can
lead to correct predictions, at least in the short term.

\section{Other Non-invertible Transformations}

The operation of calculating the mean value of several time-series is a
non-invertible transformation.

Usually it is assumed that the average value of a quantity will be a more
simple dynamics than the dynamics of the quantity itself.

Let us discuss the situation represented in Table \ref{table4}. The values of each
column are produced using the chaotic map
\begin{equation}
X_{n+1}=5.3X_{n}\pmod 1  , \label{27}%
\end{equation}
but with different initial conditions.

\begin{table}[tbp]
\begin{center}
\begin{tabular}{|c|c|c|c|c|c|c|c|c|c|c|c|} \hline
$ n $ & $ Y_n $ & \multicolumn{10}{c|}{$ X_n $} \\ \hline
0 & 0.1450 & 0.1000 & 0.1100 & 0.1200 & 0.1300 & 0.1400 & 0.1500 & 0.1600 & 0.1700 & 0.1800 & 0.1900 \\
1 & 0.6685 & 0.5300 & 0.5830 & 0.6360 & 0.6890 & 0.7420 & 0.7950 & 0.8480 & 0.9010 & 0.9540 & 0.0070 \\
2 & 0.4430 & 0.8090 & 0.0899 & 0.3708 & 0.6517 & 0.9326 & 0.2135 & 0.4944 & 0.7753 & 0.0562 & 0.0371 \\
3 & 0.4482 & 0.2877 & 0.4765 & 0.9652 & 0.4540 & 0.9428 & 0.1316 & 0.6203 & 0.1091 & 0.2979 & 0.1966 \\
4 & 0.4753 & 0.5248 & 0.5253 & 0.1158 & 0.4063 & 0.9967 & 0.6972 & 0.2877 & 0.5782 & 0.5787 & 0.0421 \\
5 & 0.4190 & 0.7815 & 0.7840 & 0.6136 & 0.1532 & 0.2827 & 0.6953 & 0.5248 & 0.0644 & 0.0669 & 0.2233 \\
6 & 0.4204 & 0.1418 & 0.1554 & 0.2519 & 0.8117 & 0.4982 & 0.6848 & 0.7813 & 0.3411 & 0.3545 & 0.1837 \\
7 & 0.6284 & 0.7515 & 0.8239 & 0.3352 & 0.3021 & 0.6404 & 0.6296 & 0.1411 & 0.8079 & 0.8788 & 0.9735 \\
8 & 0.5304 & 0.9827 & 0.3665 & 0.7767 & 0.6010 & 0.3939 & 0.3370 & 0.7476 & 0.2817 & 0.6575 & 0.1598 \\
9 & 0.5112 & 0.2084 & 0.9422 & 0.1164 & 0.1850 & 0.0875 & 0.7859 & 0.9621 & 0.4930 & 0.4846 & 0.8468 \\ \hline
\end{tabular}
\caption{The columns that correspond to $X_{n}$ are produced using the map
(\ref{27}) with different initial conditions. The variable $Y_{n}$ is the
average value of the different $X_{n}$ for a given $n$.}
\label{table4}
\end{center}
\end{table}

\begin{figure}[ptb]
\centerline{\includegraphics[width=5in]{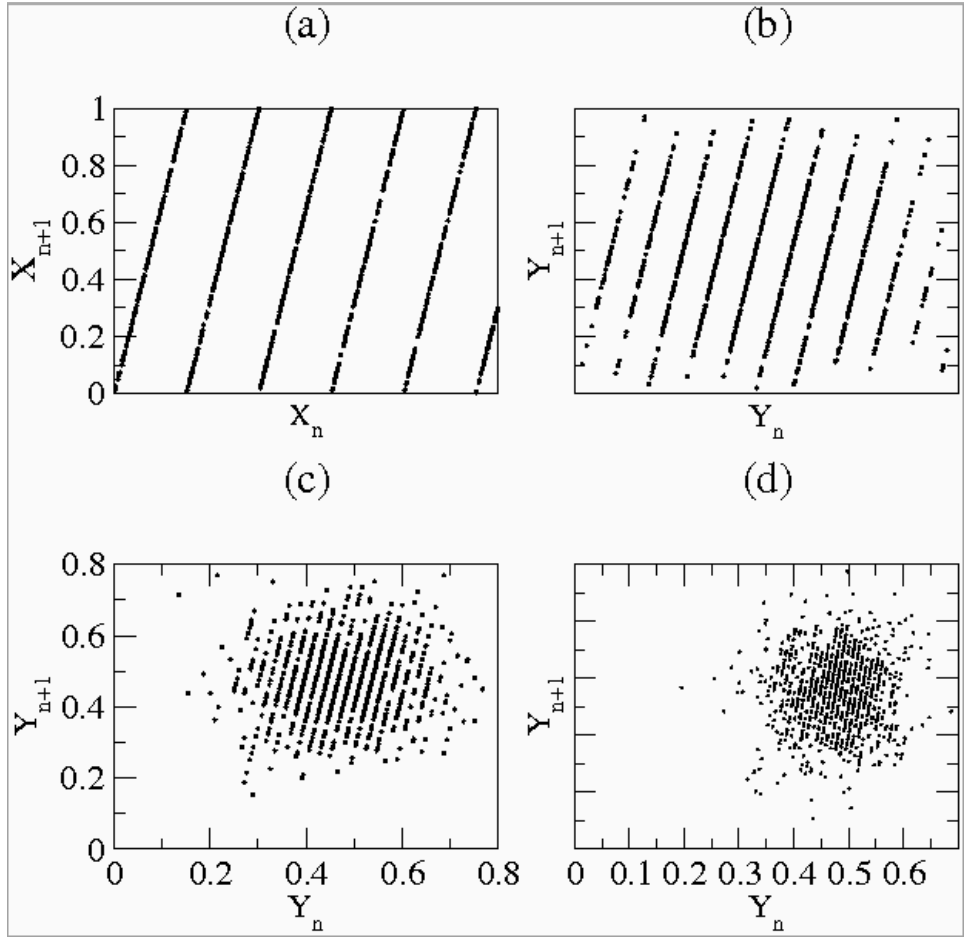}}
\caption{First return maps for $X_{n}$ as defined in Eq. (\ref{27})
and the variable $Y_{n}$, which is the average value of the different
$X_{n}$ as explained in Table \ref{table4}. (a) $N=1$. (b) $N=2$. (c) $N=8$.
(d) $N=20$.}
\label{fig15}
\end{figure}

The dynamics in each column is chaotic but predictable. This can be seen in
the first-return map shown in Fig. \ref{fig15}(a). Given a $X_{n}$, the next value is
uniquely determined.

Now let us define a new variable $Y_{n}$ as the mean value of the values
$X_{n}$ that appear in each row of the Table \ref{table4}. The result
is a time series whose complexity depends on the number of averaged columns
(see Fig. \ref{fig15}(b)-(d)). $N$ is the number of averaged column values.

\begin{figure}[ptb]
\centerline{\includegraphics[width=5in]{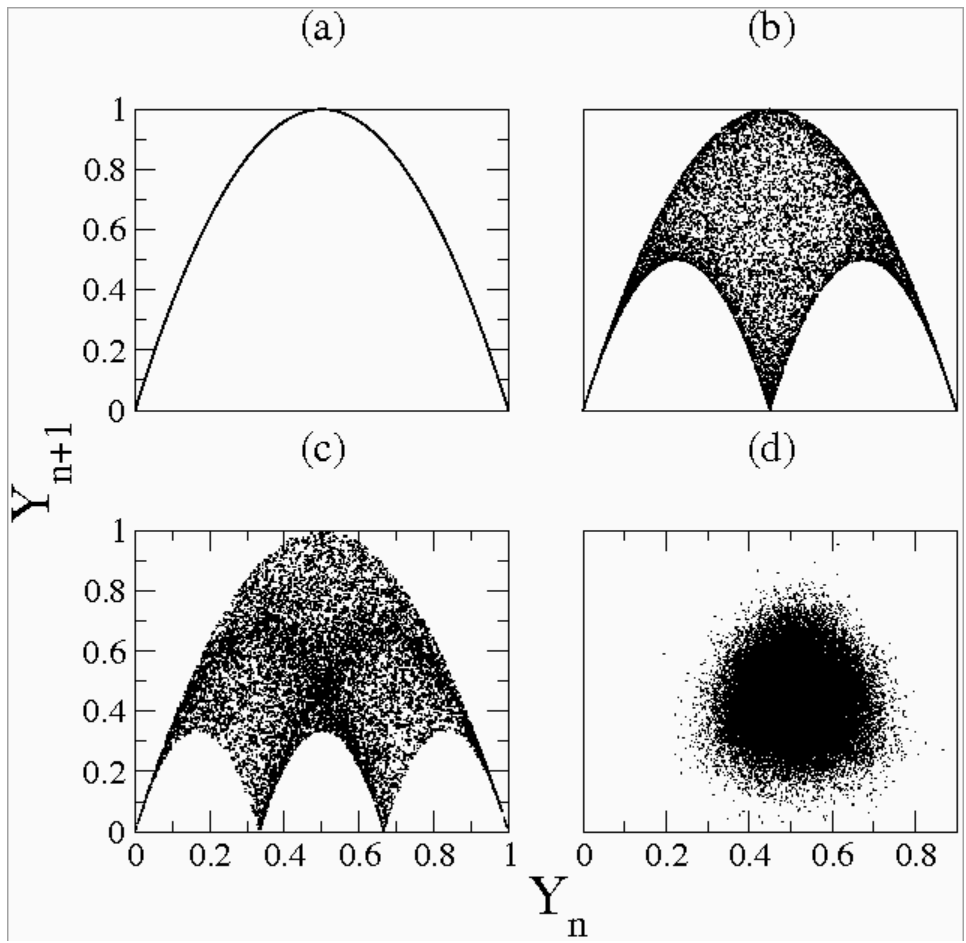}}
\caption{The same as in Fig. \ref{fig15} but the generating dynamical system
is the logistic map. (a) $N=1$. (b) $N=2$. (c) $N=3$. (d) $N=10$.}
\label{fig16}
\end{figure}

Note that for $N\rightarrow\infty$, the distribution of $Y_{n}$ tends to be
Gaussian as expected from the Central Limit Theorem.
When the chaotic map used for generating the columns is the logistic, the
dynamics of $Y_{n}$ is shown in Fig. \ref{fig16}.

It is important to remark that different forecasting methods also corroborate
that the dynamics of $Y_{n}$ becomes unpredictable
[Farmer \& Sidorowich, 1987], [Sugihara \& May, 1990].

The average operation can produce complexity also when we have only one time-series.

Suppose $X_{n}$ is a time-series produced by the chaotic map (\ref{27}). Now
define $Y_{k}$ as:
\begin{equation}
Y_{k}=\frac{1}{N}\sum\limits_{n=k}^{N+k}X_{n}. \label{28}%
\end{equation}

This dynamics\ will have the same properties as that obtained by the averaging
of several different chaotic time-series.

These results could be relevant\ to investigations of thermostatistical and
economic systems where averaging of chaotic quantities is a common practice.

The ``randomness'' found in the so-called Bernoulli random variables 
[Denker \& Woyczynski, 1998]
is also the result of the application of a non-invertible
transformation to a chaotic time-series.

This phenomenon can be seen in the following example:
\begin{equation}
Y_{n}=\phi\left(  X_{n}\right)  ,\label{29}%
\end{equation}
where $X_{n+1}=aX_{n}\pmod 1 $, $a$ is a non-integer number, $a>1$, and
$\phi\left(  t\right)  $ is defined as follows:
\begin{equation}
\phi\left(  t\right)  =\begin{cases}1, & \text{if $t\geq1/2$,}\\
0, & \text{if $t<1/2$.}
\end{cases}
\label{30}%
\end{equation}

Note that $\phi\left(  t\right)  $ is a non-invertible function.

A statistical investigation of the time-series produced by the Eq. (\ref{29})
will show that it has the same properties as the Rule 30 central column time-series.

\section{Conclusions}

We have shown that there exist dynamical systems that can generate completely
unpredictable dynamics in the sense that, given any string of generated
values, for the next outcome, there is always more than one possible value.

The mechanism responsible for the generation of randomness, in a very
general class of models and physical systems, is the presence of
non-invertible transformations of time-series that contain nonperiodic
(truncated) exponential dynamics or chaotic behavior.

Using a new theoretical framework for randomness, we have investigated some
classes of coupled map lattices. We have shown that, in some cases, these
systems can produce completely unpredictable dynamics.

Spatiotemporally chaotic systems can produce locally unpredictable dynamics
even when the global spatiotemporal dynamics is completely deterministic.
An example can be an array of coupled Josephson junctions perturbed by a
chaotic circuit like the Chua's circuit.

Local measurements of a quantity that characterizes a phenomenon in a complex
system (like the climate or the seismic events) can generate completely
unpredictable time-series. However, the global spatiotemporal data of the
phenomenon can provide the necessary information for accurate predictions, at
least in the short term.

When dealing with spatiotemporal complexity, a necessary step is an
investigation of the full space-time dynamics [Chat\'{e}, 1995]. Local probes alone are
not sufficient for efficient predictions.

Researchers need both the local and the complete spatiotemporal dynamics to
reveal the important features. The measurements should be made in an extended
zone. The more extended the zone, the better. An experimental setup should
allow the acquisition of the full space-time information. 

These results are also important in the study of economic systems
where non-invertible operations are a common practice.


\begin{thebibliography}{9}                                                                                                %

\bibitem {Bauer}Bauer, M., Heng, H. \& Martienssen, W. [1993] ``Characterization of spatiotemporal chaos from time series,''
\textit{Phys. Rev. Lett.} \textbf{71}(4), 521-524.

\bibitem {Bohr}Bohr, T., van Hecke, M., Mikkelsen, R. \& Ipsen, M. [2001] ``Breakdown of
universality in transitions to spatiotemporal chaos,''
\textit{Phys. Rev. Lett.} \textbf{86}(24), 5482-5485.

\bibitem {Brown}Brown, R. \& Chua, L. O. [1996] ``Clarifying chaos: examples and counterexamples,''
\textit{Int. J Bifurcat. Chaos} \textbf{6}(2), 219-249.

\bibitem {Bunimovich}Bunimovich, L. A. [1995] ``Coupled map lattices: one step forward and two steps back,''
\textit{Physica D} \textbf{86}(1-2), 248-255.

\bibitem {Chate1}Chat\'{e}, H. \& Manneville, P. [1988] ``Spatio-temporal intermittency in coupled map lattices,''
\textit{Physica D} \textbf{32}(3), 409-422.

\bibitem {Chate2}Chat\'{e}, H. [1995] ``On the analysis of spatiotemporally chaotic data,''
\textit{Physica D} \textbf{86}(1-2), 238-247.

\bibitem {Chua}Chua, L. O., Desoer, C. A. \& Kuh, E. S. [1987] \textit{Linear and Nonlinear Circuits}
(McGraw-Hill, New York).

\bibitem {Comte}Comte, J. C. \& Marqui\'{e}, P. [2002] ``Generation of nonlinear current-voltage characteristics: a general method,''
\textit{Int. J Bifurcat. Chaos} \textbf{12}(2), 447-449.

\bibitem {Crutchfield}Crutchfield, J. P. \& Kaneko, K. [1988] ``Are attractors relevant to turbulence?,''
\textit{Phys. Rev. Lett.} \textbf{60}(26), 2715-2718.

\bibitem {Denker}Denker, M. \& Woyczynski, W. A. [1998] \textit{Introductory Statistics and Random Phenomena}
(Birkh\"{a}user, Berlin).

\bibitem {Farmer}Farmer, J. D. \& Sidorowich, J. J. [1987] ``Predicting chaotic time series,''
\textit{Phys. Rev. Lett.} \textbf{59}(8), 845-848.

\bibitem {Glass}Glass, L. \& Mackey, M. C. [1998] \textit{From Clocks to Chaos, The Rhythms of Life}
(Princeton University Press, Princeton).

\bibitem {Gonzalez1}Gonz\'{a}lez, J. A., Guerrero, L. E. \& Bellor\'{\i}n, A. [1996] ``Self-excited soliton motion,''
\textit{Phys. Rev. E} \textbf{54}(2), 1265-1273.

\bibitem {Gonzalez2}Gonz\'{a}lez, J. A., Mello,  B. A., Reyes, L. I. \& Guerrero, L. E. [1998]
``Resonance phenomena of a solitonlike extended object in a bistable potential,'' \textit{Phys. Rev. Lett.} \textbf{80}(7), 1361-1364.

\bibitem {Gonzalez3}Gonz\'{a}lez, J. A., Mart\'{\i}n-Landrove, M. \& Trujillo, L. [2000] ``Absolutely unpredictable chaotic sequences,''
\textit{Int. J Bifurcat. Chaos} \textbf{10}(8), 1867-1874.

\bibitem {Gonzalez4}Gonz\'{a}lez, J. A., Reyes, L. I., Su\'{a}rez, J. J., Guerrero, L. E. \& Guti\'{e}rrez, G. [2002]
``A mechanism for randomness,'' \textit{Phys. Lett. A} \textbf{295}(1), 25-34.

\bibitem {Grassberger}Grassberger, P. \& Scheiber, T. [1991] ``Phase transitions in coupled map lattices,''
\textit{Physica D} \textbf{50}(2), 177-188.

\bibitem {Grigoriev1}Grigoriev, R. O., Cross, M. C. \& Schuster, H. G. [1997] ``Pinning control of spatiotemporal chaos,''
\textit{Phys. Rev. Lett.} \textbf{79}(15), 2795-2798.

\bibitem {Grigoriev2}Grigoriev, R. O. \& Schuster, H. G. [1998] ``Solvable model for spatiotemporal chaos,''
\textit{Phys. Rev. E} \textbf{57}(1), 388-396.

\bibitem {Guerrero}Guerrero, L. E., Bellor\'{\i}n, A., Carb\'{o}, J. R. \& Gonz\'{a}lez, J. A. [1999]
``Spatiotemporal chaotic dynamics of solitons with internal structure in the presence of finite-width inhomogeneities,''
\textit{Chaos, Solitons \& Fractals} \textbf{10}(9), 1491-1512.

\bibitem {Hansel}Hansel, D. \& Sompolinsky, H. [1993] ``Solvable model of spatiotemporal chaos,''
\textit{Phys. Rev. Lett.} \textbf{71}(17), 2710-2713.

\bibitem {Hastings}Hastings, M. B., Olson Reichhardt, C. J. \& Reichhardt, C. [2003]
``Ratchet cellular automata,'' \textit{Phys. Rev. Lett.} \textbf{90}(24), Art. No. 247004.

\bibitem {Israeli}Israeli, N. \& Goldenfeld, N. [2004] ``Computational irreducibility and the predictability of complex physical systems,''
\textit{Phys. Rev. Lett.} \textbf{92}(7), Art. No. 074105.

\bibitem {Jackson} Jackson, A. [1991] \textit{Perspectives of Nonlinear Dynamics}
(Cambridge University Press, New York).

\bibitem {Kaneko1}Kaneko, K. [1985] ``Spatiotemporal intermittency in coupled map lattices,''
\textit{Prog. Theor. Phys.} \textbf{74}(5), 1033-1044.

\bibitem {Kaneko2}Kaneko, K. [1989] ``Spatiotemporal chaos in one-dimensional and two-dimensional coupled map lattices,''
\textit{Physica D} \textbf{37}(1-3), 60-82.

\bibitem {Kaneko3}Kaneko, K. \& Konishi, T. [1989] ``Diffusion in Hamiltonian dynamical systems with many degrees of freedom,''
\textit{Phys. Rev. A} \textbf{40}(10), 6130-6133.

\bibitem {Kaneko4}Kaneko, K. [1990] ``Globally coupled chaos violates the law of large numbers but not the central-limit theorem,''
\textit{Phys. Rev. Lett.} \textbf{65}(12), 1391-1394.

\bibitem {Kaneko5}Kaneko, K. [1990a] ``Supertransients, spatiotemporal intermittency and stability of fully developed spatiotemporal chaos,''
\textit{Phys. Lett. A} \textbf{149}(2-3), 105-112.

\bibitem {Kaneko6}Kaneko, K. [1992] ``Overview of coupled map lattices,''
\textit{Chaos} \textbf{2}(3), 279-282.

\bibitem {Kaneko7}Kaneko, K. \& Tsuda, I. [2003] ``Chaotic itinerancy,''
\textit{Chaos} \textbf{13}(3), 926-936.

\bibitem {Kantz}Kantz, H. \& Schreiber, T. [1997] \textit{Nonlinear Time Series Analysis}
(Cambridge University Press, Cambridge, U.K.).

\bibitem {Kuramoto}Kuramoto, Y. [1984] \textit{Chemical Oscillations, Waves, and Turbulence}
(Springer, Berlin).

\bibitem {Lorenz}Lorenz, E. N. [1993] \textit{The Essence of Chaos}
(University of Washington Press, New York).

\bibitem {Matsumoto1}Matsumoto, T., Chua, L. O. \& Komuro, M. [1985] ``The double scroll,''
\textit{IEEE Trans. Circuits Syst.} \textbf{32}(8), 797-818.

\bibitem {Matsumoto2}Matsumoto, T., Chua, L. O. \& Komuro, M. [1987] ``Birth and death of the double scroll,''
\textit{Physica D} \textbf{24}(1-3), 97-124.

\bibitem {MayerKress}Mayer-Kress, G. \& Kaneko, K. [1989] ``Spatiotemporal chaos and noise,''
\textit{J. Stat. Phys.} \textbf{54}(5-6), 1489-1508.

\bibitem {Moon}Moon, F. C. [1991] \textit{Chaotic and Fractal Dynamics. An Introduction for Applied Scientists and Engineers}
(Wiley, New York).

\bibitem {Pikovsky}Pikovsky, A. S. \& Kurths, J. [1994] ``Do globally coupled maps really violate the law of large numbers?,''
\textit{Phys. Rev. Lett.} \textbf{72}(11), 1644-1646.

\bibitem {Politi}Politi, A. \& Torcini, A. [1992] ``Towards a statistical mechanics of spatiotemporal chaos,''
\textit{Phys. Rev. Lett.} \textbf{69}(24), 3421-3424.

\bibitem {Schuster}Schuster, H. G. [1995] \textit{Deterministic Chaos. An Introduction}
(Wiley, New York).

\bibitem {Shibata}Shibata, T. \& Kaneko, K. [1998] ``Collective chaos,''
\textit{Phys. Rev. Lett.} \textbf{81}(19), 4116-4119.

\bibitem {Strogatz}Strogatz, S. H. [1994] 
\textit{Nonlinear dynamics and Chaos: With Applications in Physics, Biology, Chemistry and Engineering}
(Addison-Wesley, Reading (MA)).

\bibitem {Sugihara}Sugihara, G. \& May, R. M. [1990] ``Nonlinear forecasting as a way of distinguishing chaos from measurement
error in time series,'' \textit{Nature} \textbf{344}(6268), 734-741.

\bibitem {vonNeumann}von Neumann, J. \& Burks, A. W. [1966] \textit{Theory of Self-Reproducing Automata}
(University of Illinois Press, Urbana).

\bibitem {Wackerbauer}Wackerbauer, R. \& Showalter, K. [2003] ``Collapse of spatiotemporal chaos,''
\textit{Phys. Rev. Lett.} \textbf{91}(17), Art. No. 174103.

\bibitem {Willeboordse}Willeboordse, F. T. [2003] ``The spatial logistic map as a simple prototype for spatiotemporal chaos,''
\textit{Chaos} \textbf{13}(2), 533-540.

\bibitem {Wolf}Wolf, A., Swift, J. B., Swinney, H. L. \& Vastano, J. A. [1985] ``Determining Lyapunov exponents
from a time series,'' \textit{Physica D} \textbf{16}(3), 285-317. 

\bibitem {Wolfram1}Wolfram, S. [1983] ``Statistical mechanics of cellular automata,''
\textit{Rev. Mod. Phys.} \textbf{55}(3), 601-644.

\bibitem {Wolfram2}Wolfram, S. [1984] ``Cellular automata as models of complexity,''
\textit{Nature} \textbf{311}(5985), 419-424.

\bibitem {Wolfram3}Wolfram, S. [1984a] ``Universality and complexity in cellular automata,''
\textit{Physica D} \textbf{10}(1-2), 1-35.

\bibitem {Wolfram4}Wolfram, S. [1986] ``Random sequence generation by cellular automata,''
\textit{Adv. Appl. Math.} \textbf{7}(2), 123-169.

\end{thebibliography}
\end{document}